\documentclass[prd,aps,nofootinbib,floatfix,10 pt]{revtex4}
\usepackage{amssymb}
\usepackage{amsmath,graphicx,color,epsfig}

\begin{document}

\title{Dirac equation for quasi-particles in graphene in an external electromagnetic field and chiral anomaly}
\author{Riazuddin}
\email{riazuddin@ncp.edu.pk}
\affiliation{National Centre for Physics, Quaid-i-Azam University Campus, 45320
Islamabad, Pakistan}
\date{\today}

\begin{abstract}
There is evidence for existence of massless Dirac quasi-particles in
graphene, which satisfy Dirac equation in (1+2) dimensions near the so
called Dirac points which lie at the corners at the graphene's brilluoin
zone. It is shown that parity operator in (1+2) dimensions play an
interesting role and can be used for defining conserved chiral currents
[there is no $\gamma ^{5}$ in (1+2) dimensions]. It is shown that the
"anomalous" current induced by an external gauge field can be related to the
anomalous divergence of an axial vector current which arises due to quantum
radiative corrections provided by triangular loop Feynman diagrams in
analogy with the corresponding axial anomaly in (1+3) dimensions.
\end{abstract}

\maketitle

Recent progress in the experimental realization of a single layer problem of
graphene has lead to extensive exploration of electronic properties in this
system. Experimental and theoretical studies have shown that the nature of
quasiparticles in these two-dimensional system are very different from those
of the conventional two-dimensional electron gas (2DEG) system realized in
the semiconductor heterostructures. Graphene has a honeycomb lattice of
carbon atoms. The quasiparticles in graphene have a band structure in which
electron and hole bands touch at two points in the Brillouin zone. At these
Dirac points, the quasiparticles obey the massless Dirac equation in (1+2)
dimensions \cite{1}. In other words, they behave as massless Dirac fermions
leading to a linear dispersion relation $\epsilon _{k}=vk$ (with the
characteristic velocity $v\simeq 10^{6}$m/s). This difference in the nature
of the quasiparticles in graphene from conventional 2DEG has given rise to a
host of new and unusual phenomena such as anomalous quantum Hall effects and
a $\pi $ Berry phase \cite{1, 2}. These transport experiments have shown
results in agreement with the presence of Dirac fermions. The 2D Dirac-like
spectrum was confirmed recently by cyclotron resonance measurements and also
by angle resolved photoelectron spectroscopy (ARPEC) measurements in
monolayer graphene \cite{3}. Recent theoretical work on graphene multilayer
has also shown the existence of Dirac electrons with a linear energy
spectrum in monolayer graphene \cite{4}.

The Dirac points lie at the corners of the graphene's Brillouin zone and
have their position vectors in momentum space as \cite{5}.
\begin{equation*}
\mathbf{K}=\frac{2\pi }{3a}(1,\sqrt{3}),\text{ \qquad }\mathbf{K}^{\prime }=%
\frac{2\pi }{3a}(1,-\sqrt{3})
\end{equation*}%
where $a$ is the carbon-carbon distance.

Near the Dirac point $\mathbf{K}^{^{\prime }}$, the Dirac equation takes the
covariant form \cite{6}
\begin{equation}
i(\gamma ^{\mu }\partial _{\mu })\psi =0  \label{1}
\end{equation}%
where
\begin{equation*}
\partial _{0}=\frac{1}{v_{f}}\frac{\partial }{\partial t}
\end{equation*}%
and%
\begin{equation}
\gamma ^{\mu }\partial _{\mu }=\gamma ^{0}\partial _{0}+\gamma ^{1}\partial
_{1}+\gamma ^{2}\partial _{2}  \label{2}
\end{equation}

Now it is known \cite{7} that in 3 space-time dimensions there
exists two inequivalent representations for $\gamma $-matrices [this
is true for any odd number of space-time dimensions]:
\begin{eqnarray}
\gamma ^{0} &=&\sigma ^{3},\text{ }\gamma ^{1}=i\sigma ^{1},\gamma
^{2}=i\sigma ^{2}  \notag \\
\gamma ^{0} &=&\sigma ^{3},\text{ }\gamma ^{1}=i\sigma ^{1},\gamma
^{2}=-i\sigma ^{2}  \label{3}
\end{eqnarray}

One can take the second representation for the Dirac equation near the Dirac
point K, which is obtained from K$^{^{\prime }}$ by the parity operation:
\begin{equation*}
x^{1}\longleftrightarrow x^{1}\text{, }x^{2}\longleftrightarrow -x^{2}\text{
}
\end{equation*}

Taking the two representations mentioned above into account one can write
the parity conserving Lagrangian as
\begin{equation*}
\mathcal{L}=\overline{\psi }_{+}(i\partial )\psi _{+}+\overline{\psi }_{-}(i%
\widetilde{\partial })\psi _{-}
\end{equation*}%
where
\begin{eqnarray}
\partial &=&\gamma ^{0}\partial _{0}+\gamma ^{1}\partial _{1}+\gamma
^{2}\partial _{2}  \notag \\
\widetilde{\partial } &=&\gamma ^{0}\partial _{0}+\gamma ^{1}\partial
_{1}-\gamma ^{2}\partial _{2}  \label{4}
\end{eqnarray}%
Parity operation takes the solutions in one representation to the other:
\begin{eqnarray}
\psi _{+}^{p}(x^{p}) &=&-\eta _{p}\psi _{-}(x)  \notag \\
\psi _{-}^{p}(x^{p}) &=&-\eta _{p}\psi _{+}(x)  \label{5}
\end{eqnarray}%
where $x^{p}=(x^{0},x^{1},-x^{2})$. It is convenient to transform to new
fields \cite{7}
\begin{eqnarray}
\psi _{A} &=&\psi _{+}  \notag \\
\psi _{B} &=&i\gamma ^{2}\psi _{-}  \label{6}
\end{eqnarray}%
The Lagrangian (\ref{4}) can then be written as
\begin{equation}
\mathcal{L}=\overline{\psi }_{A}(i\gamma ^{\mu }\partial _{\mu })\psi _{A}+%
\overline{\psi }_{B}(i\gamma ^{\mu }\partial _{\mu })\psi _{B}  \label{7}
\end{equation}%
It is instructive to \ put mass term in the Lagrangian (\ref{7}), which one
can always put equal to zero:%
\begin{equation}
\mathcal{L}=\overline{\psi }_{A}(i\gamma ^{\mu }\partial _{\mu })\psi _{A}+%
\overline{\psi }_{B}(i\gamma ^{\mu }\partial _{\mu })\psi _{B}-mv_{f}(%
\overline{\psi }_{A}\psi _{A}-\overline{\psi }_{B}\psi _{B})  \label{8}
\end{equation}%
where under the parity operation%
\begin{equation}
\psi _{A,B}^{P}(x^{P})=\eta _{P}\sigma ^{2}\psi _{B,A}(x^{P})  \label{9}
\end{equation}

The Hamiltonian density is%
\begin{equation}
\mathcal{H}=v_{f}[\overline{\psi }_{A}(-i\gamma ^{i}\partial _{i})\psi _{A}+%
\overline{\psi }_{B}(-i\gamma ^{i}\partial _{i})\psi _{B}+mv_{f}(\overline{%
\psi }_{A}\psi _{A}-\overline{\psi }_{B}\psi _{B})]  \label{10}
\end{equation}%
The Hamiltonian density has the so called conjugate symmetry \cite{8}, $\psi
_{A}\leftrightarrow \sigma ^{3}\psi _{B},$ in the sense that $\mathcal{%
H\rightarrow -H}$.

It may be noted that the Lagrangian (\ref{8}) is invariant, even in the
presence of the mass term, two independent transformations%
\begin{equation}
\psi _{A}\rightarrow e^{i\alpha _{A}}\psi _{A},\text{ }\psi _{B}\rightarrow
e^{i\alpha _{B}}\psi _{B}  \label{11}
\end{equation}%
where $\alpha _{A}$ and $\alpha _{B}$ are real, and has thus $%
U_{A}(1)\otimes U_{B}(1)$ symmetry. The corresponding conserved currents are
\begin{equation}
J_{A}^{\mu }=\overline{\psi }_{A}\gamma ^{\mu }\psi _{A},\text{ }J_{B}^{\mu
}=\overline{\psi }_{B}\gamma ^{\mu }\psi _{B}  \label{12}
\end{equation}%
One can form even (odd) combination corresponding to "vector" ("axial
vector") under parity%
\begin{equation}
J_{\pm }^{\mu }=1/2[\overline{\psi }_{A}\gamma ^{\mu }\psi _{A}\pm \overline{%
\psi }_{B}\gamma ^{\mu }\psi _{B}]  \label{13}
\end{equation}

In (1+2) dimensions, there is no $\gamma ^{5}$ available as in (1+3)
dimensions. But still one may generate "chiral" currents. In fact under the
infinitesimal transformations \cite{7}.%
\begin{equation*}
\psi _{A,B}\rightarrow \psi _{A,B}^{\prime }=\text{ }\psi _{A,B}^{\prime
}+i\alpha \psi _{B,A}
\end{equation*}%
and%
\begin{equation*}
\psi _{A,B}\rightarrow \psi _{A,B}^{\prime }=\text{ }\psi _{A,B}^{{}}\pm
\alpha \psi _{B,A}
\end{equation*}%
the Lagrangian (\ref{8}) respectively transforms into%
\begin{eqnarray}
\mathcal{L}\mathcal{\rightarrow L}_{1}=\mathcal{L-}%
2imv_{f}\alpha (\overline{\psi }_{A}\psi _{B}-\overline{\psi }_{B}\psi _{A})
\notag \\
\mathcal{L}\mathcal{\rightarrow L}_{2}=\mathcal{L-}2mv_{f}(%
\overline{\psi }_{A}\psi _{B}+\overline{\psi }_{B}\psi _{A})  \label{14}
\end{eqnarray}%
The corresponding conserved "chiral" currents in the absence of mass are%
\begin{eqnarray}
J_{3}^{\mu } &=&\frac{1}{2}(\overline{\psi }_{A}\gamma ^{\mu }\psi _{B}+%
\overline{\psi }_{B}\gamma ^{\mu }\psi _{A})  \label{15a} \\
J_{5}^{\mu } &=&-i\frac{1}{2}(\overline{\psi }_{A}\gamma ^{\mu }\psi _{B}-%
\overline{\psi }_{B}\gamma ^{\mu }\psi _{A})  \label{15b}
\end{eqnarray}%
These currents respectively correspond to "vector" and "axial vector" under
parity. In the presence of mass term in the Lagrangian (\ref{8})%
\begin{eqnarray}
\partial _{\mu }J_{3}^{\mu } &=&imv_{f}(\overline{\psi }_{A}\psi _{B}-%
\overline{\psi }_{B}\psi _{A})  \label{16a} \\
\partial _{\mu }J_{5}^{\mu } &=&mv_{f}(\overline{\psi }_{A}\psi _{B}+%
\overline{\psi }_{B}\psi _{A})  \label{16b}
\end{eqnarray}

An external gauge field $A_{\mu }$ (electromagnetic) can be introduced by
replacing the ordinary derivative by the covariant derivative%
\begin{equation}
\partial _{\mu }\rightarrow D_{\mu }=\partial _{\mu }+\frac{ie}{c}A_{\mu }
\label{17}
\end{equation}

This gives the Dirac equation in (1+2) dimensions%
\begin{equation}
\lbrack i\gamma ^{\mu }D_{\mu }\mp mv_{f}]\psi _{A,\text{ }B}=0,\text{ }
\label{18}
\end{equation}%
By multiply on the left by $(-i\gamma ^{\nu }D_{\nu }\mp mv_{f})$
one can put the
resulting equation in the Pauli form%
\begin{equation}
\lbrack D^{\mu }D_{\mu }+\frac{e}{2c}\sigma ^{\mu \nu }F_{\mu \nu
}+m^{2}v_{f}^{2}]\psi _{A,\text{ }B}=0  \label{19}
\end{equation}%
This equation differs from the Klein Gorden equation in the term $\frac{e}{c}%
\sigma ^{\mu \nu }F_{\mu \nu },$ $F_{\mu \nu }=\partial _{\mu }A_{\upsilon
}-\partial _{\nu }A_{\mu }.$ Using
\begin{equation}
\sigma ^{\mu \nu }=\epsilon ^{\lambda \mu \nu }\gamma _{\lambda }
\label{20a}
\end{equation}%
one can write%
\begin{equation}
\frac{e}{2c}\sigma ^{\mu \nu }F_{\mu \nu }=1/2f^{\lambda }\gamma
_{\lambda } \label{20b}
\end{equation}%
where $f^{\lambda }$ is current induced by the external gauge field $A_{\mu
} $ \cite{8, 9}$.$%
\begin{equation}
f^{\lambda }=\frac{e}{c}\epsilon ^{\lambda \mu \nu }F_{\mu \nu }
\label{21}
\end{equation}%
and has abnormal parity. The corresponding induced charge is
\begin{equation}
Q=\int d^{3}x\text{ }f^{0}(x)  \label{22}
\end{equation}%
where%
\begin{equation}
f^{0}(x)=\frac{e}{c}\epsilon ^{ij}F_{ij}=2\frac{e}{c}(\vec{\nabla}\times
\vec{A})_{3}=2\frac{e}{c}B  \label{23}
\end{equation}%
Here $B$ is the magnetic field perpendicular to the $x-y$ plane. Thus $Q=%
\frac{e}{c}\Phi $, where $\Phi $ is the magnetic flux.

The parity conserving Lagrangian which gives Eq. (21) is%
\begin{equation*}
\mathcal{L}_{\pm }=[\overline{\psi }_{A}(D^{\mu }D_{\mu
}+m^{2}v_{f}^{2})\psi _{A}-\overline{\psi }_{B}(D^{\mu }D_{\mu
}+m^{2}v_{f}^{2})\psi _{B}]+1/2[(\overline{\psi }_{A}\gamma ^{\mu
}\psi _{A}-\overline{\psi }_{B}\gamma ^{\mu }\psi _{B})]f_{\mu }
\end{equation*}%
so that $f_{\mu }$ is coupled with the current $J_{-}^{\mu }$ given in Eq.
(13) \cite{8}. The Lagrangian is invariant under $U_{A}(1)\otimes U_{B}(1).$

Next we discuss whether $f_{\mu }$ can be related to "anomalous" divergence
of some axial current, which arises due to quantum corrections. Obviously
such a current can not be $J_{-}^{\mu }$. Then the remaining axial current
is $J_{5}^{\mu }$ given in Eq. (16b). Here analogy with the axial vector
"anomalous" divergence in (1+3) dimensions, namely \cite{10}%
\begin{eqnarray}
\partial _{\mu }J_{5}^{\mu } &=&-\frac{e^{2}}{32\pi ^{2}}\epsilon ^{\mu \nu
\rho \sigma }F_{\mu \nu }F_{\rho \sigma }  \notag \\
&=&\frac{e^{2}}{4\pi ^{2}}\overrightarrow{E}.\overrightarrow{B}  \label{25}
\end{eqnarray}%
is useful. As is well known this divergence arises from quantum corrections
provided by triangle graph which has two vector vertices and one axial
vector vertex or that provided by its divergence%
\begin{equation*}
\partial _{\mu }J_{5}^{\mu }=mc(\overline{u}i\gamma _{5}u-\overline{d}%
i\gamma _{5}d)
\end{equation*}%
where $u$ and $d$ denote up and down quarks ($m$ is the quark mass), which
provide the internal legs of the triangle. Note that although quark mass $m$
appears above, but Eq. (27) is independent of quark mass. In our case we
have corresponding $\psi _{A}$ and $\psi _{B}$ fields, which appear in the
Lagrangian (8) or Hamiltonian (10). Noting that $\partial _{\mu }J_{5}^{\mu
} $ involves $m\overline{\psi }_{A}\psi _{B}$\ and $m\overline{\psi }%
_{B}\psi _{A},$ the relevant Feynman graphs are shown in Fig 1

\begin{figure}[th]
\includegraphics[scale=0.6]{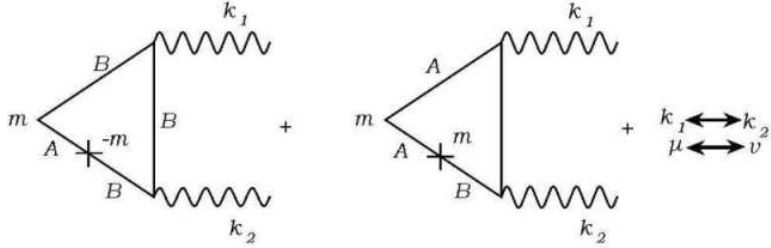}
\caption{Triangle diagrams for "anomalous" current divergence.}
\end{figure}

Note that it is essential to put mass transitions shown (see analogy with
Majorana neutrinos), such mass transitions are provided by the divergence of $%
\partial _{\mu }J_{3}^{\mu }$\ given in Eq. (17). Noting that A and B
propagators involve opposite masses, the matrix elements are given by%
\begin{eqnarray}
T_{\mu \nu } &=&mv_{f}e^{2}\int \frac{d^{D}l}{(2\pi )^{D}}\{Tr[\frac{1}{{%
\not {l}+k_{1}-mv_{f}}}\gamma _{\mu }\frac{1}{\not{l}-mv_{f}}\gamma _{\nu }(%
\frac{1}{\not{l}-\not{k_{2}}-mv_{f}}mv_{f}\frac{1}{\not{l}-k_{2}+mv_{f}})
\notag \\
&&+\frac{1}{\not{l}+k_{1}+mv_{f}}\gamma _{\mu }\frac{1}{\not{l}+mv_{f}}%
\gamma _{\nu }(\frac{1}{\not{l}-\not{k_{2}}+mv_{f}}(-mv_{f})\frac{1}{\not{l}-%
\not{k_{2}}-mv_{f}})]\}  \notag \\
&=&m^{2}v_{f}^{2}e^{2}\int \frac{d^{D}l}{(2\pi )^{D}}\frac{Tr[(\not{l}-\not{%
k_{1}}+mv_{f})\gamma ^{\mu }(\not{l}+mv_{f})\gamma ^{\nu }-(\not{l}+\not{%
k_{1}}-mv_{f})\gamma ^{\mu }(\not{l}-mv_{f})\gamma ^{\nu }]}{%
[(l+k_{1})^{2}-m^{2}v_{f}^{2}][l^{2}-m^{2}v_{f}^{2}][(l-k_{2})^{2}-m^{2}v_{f}^{2}]%
}+\left.
\begin{tabular}{l}
$k_{1}\leftrightarrow k_{2}$ \\
$\mu \leftrightarrow \nu $%
\end{tabular}%
\right.  \label{26}
\end{eqnarray}%
The numerator in the integral which contributes is,
\begin{eqnarray}
N^{\mu } &=&2mv_{f}\text{ }Tr[\gamma ^{\rho }\gamma ^{\mu }\gamma ^{\nu
}(l+k_{1})_{\rho }+\gamma ^{\rho }\gamma ^{\mu }\gamma ^{\nu }l_{\rho }]
\notag \\
&=&-4mv_{f}i[\epsilon ^{\rho \mu \nu }(l+k_{1})_{\rho }+\epsilon ^{\mu \rho
\nu }l_{\rho }]  \notag \\
&&-4mv_{f}i\epsilon ^{\mu \nu \rho }k_{1\rho }  \label{27}
\end{eqnarray}

Using the Feynman parametrization, the denominator takes the the form,
\begin{equation*}
\lbrack (l+k_{1}x-k_{2}y)^{2}-\Delta ]^{3}
\end{equation*}%
where
\begin{equation*}
\Delta =m^{2}v_{f}^{2}-xk_{1}^{2}-yk_{2}^{2}+(xk_{1}-yk_{2})^{2}
\end{equation*}%
Making the shift $l\rightarrow l-(k_{1}x-k_{2}y)$, the denominator becomes $%
(l^{2}-\Delta )^{3}$ and the dimensional regularization gives [D=3]%
\begin{eqnarray}
T_{\mu \nu } &=&-4m^{3}v_{f}^{3}e^{2}i\epsilon ^{\mu \nu \rho }k_{1\rho
}\int_{0}^{1}dx\int_{0}^{1-x}dy\frac{(-1)^{3}i}{(4\pi )^{d/2}}\frac{\Gamma
(3-D/2)}{\Gamma (3)}(\frac{1}{\Delta })^{3-D/2}  \notag \\
&&+\left.
\begin{tabular}{l}
$k_{1}\leftrightarrow k_{2}$ \\
$\mu \leftrightarrow \nu $%
\end{tabular}%
\right.  \label{28}
\end{eqnarray}

For photons on the mass shell%
\begin{equation*}
k_{1}^{2}=0,\text{ \ \ \ \ }k_{2}^{2}=0
\end{equation*}%
and putting $2k_{1}.k_{2}=q^{2},$ $\Delta =m^{2}v_{f}^{2}-q^{2}xy$. Thus we
obtain, neglecting terms of order $q^{2}.$%
\begin{eqnarray}
T_{\mu \nu } &=&\frac{e^{2}}{16\pi }\frac{m^{3}}{(m^{6})^{1/2}}\epsilon
^{\mu \nu \rho }k_{1\rho }+\left.
\begin{tabular}{l}
$k_{1}\leftrightarrow k_{2}$ \\
$\mu \leftrightarrow \nu $%
\end{tabular}%
\right.  \notag \\
&=&\frac{e^{2}}{16\pi }(\text{sign }m)\epsilon ^{\mu \nu \rho
}(k_{1}-k_{2})_{\rho }  \label{29}
\end{eqnarray}

Thus finally%
\begin{equation}
\partial _{\mu }J_{5}^{\mu }=\frac{e^{2}}{16\pi }(\text{sign }m)\varepsilon
_{\mu }(k_{1})\varepsilon _{\nu }(k_{2})\epsilon ^{\mu \nu \rho
}(k_{1}-k_{2})_{\rho }  \label{30}
\end{equation}

It may be noted while in (1+3) dimensions $\partial _{\mu }J_{5}^{\mu }$ is
independent of $m$; in (1+2) dimensions, the corresponding quantity $%
\partial _{\mu }J_{5}^{\mu }$ is also independent of the magnitude of $m$
but does depend on its sign (which is typical for odd space-time
dimensions). In both cases mass was used as a regulator. In configuration
space Eq. (32) takes the form%
\begin{equation}
\partial _{\mu }J_{5}^{\mu }=\frac{e^{2}}{16\pi }(\text{sign }m)A_{\lambda
}f^{\lambda }=\frac{e^{2}}{16\pi }(\text{sign }m)\epsilon ^{\mu \nu \lambda
}F_{\mu \nu }A_{\lambda }  \label{31}
\end{equation}%
[comparing with Eq. (27) in (1+3) dimensions]. Now in terms of electric and
magnetic fields.%
\begin{equation*}
\epsilon ^{\mu \nu \lambda }F_{\mu \nu }A_{\lambda }=-2A^{0}B^{3}-2(%
\overrightarrow{A}\times \overrightarrow{E})^{3}=-2B[A_{0}+\overrightarrow{E}%
.\overrightarrow{r}]
\end{equation*}%
where $B$ is along $z-$axis and we have used $\overrightarrow{A}=\frac{1}{2}%
\overrightarrow{B}\times \overrightarrow{r}$ so that $\overrightarrow{B}=%
\overrightarrow{\triangledown }\times \overrightarrow{A}$. One can select a
gauge in which $A_{0}=0$ then Eq. (31) becomes%
\begin{equation}
\partial _{\mu }J_{5}^{\mu }=-\frac{e^{2}}{8\pi }(\text{sign }m)B(%
\overrightarrow{E}.\overrightarrow{r})  \label{32}
\end{equation}%
[compare with second line ($\overrightarrow{E}.\overrightarrow{B}$) of Eq.
(27) for axial anomaly in (1+3)]. It is instructive to also calculate $%
\partial _{\mu }J_{3}^{\mu }$ from the $\Delta -$graph, the only change one
has to make is to change the overall sign of the second term in Eq. (\ref{26}%
) and $m$ multiplying the integral to $im.$ This changes $N^{\mu }$ in Eq. (\ref{27}%
) to

\begin{eqnarray}
N^{\mu } &=&2\text{ }Tr[\gamma ^{\rho }\gamma ^{\mu }\gamma ^{\sigma }\gamma
^{\nu }(l+k_{1})^{\rho }l^{\sigma }+m^{2}v_{f}^{2}\gamma ^{\mu }\gamma ^{\nu
}]  \notag \\
&=&4[(l+k_{1})^{\mu }l^{\nu }+(l+k_{1})^{\nu }l^{\mu }-(l+k_{1})l\text{ }%
g^{\mu \nu }+m^{2}v_{f}^{2}g^{\mu \nu }]  \label{33}
\end{eqnarray}%
After making the shift $l\rightarrow l-(k_{1}x-k_{2}y)$ and noting that
terms linear in $l$ do not contribute, one obtains%
\begin{eqnarray}
T_{\mu \nu } &=&4im^{2}v_{f}^{2}\left\{\frac{i}{(4\pi )^{d/2}}%
\frac{1}{\Gamma (3)} \int_{0}^{1} dx\int_{0}^{1-x}dy\text{ }g^{\mu \nu }\frac{1-d}{2}%
\Gamma (2-d/2)(\frac{1}{\Delta })^{2-d/2}+k_{2}^{\mu }k_{1}^{\nu }\text{ }xy%
\text{ }\Gamma (3-d/2)(\frac{1}{\Delta })^{3-d/2}\right\}  \notag \\
&&+\left.
\begin{tabular}{l}
$k_{1}\leftrightarrow k_{2}$ \\
$\mu \leftrightarrow \nu $%
\end{tabular}%
\right.  \label{34}
\end{eqnarray}%
where we have put $k_{1}^{2}=0=k_{2}^{2},$ $k_{1}.\epsilon
_{1}=k_{2}.\epsilon _{2}$=0 and then neglecting
$2k_{1}.k_{2}=q^{2},$ so that $\Delta =m^{2}v_{f}^{2}.$ Then finally
\begin{equation}
\partial _{\mu }J_{3}^{\mu }=\frac{e^{2}}{4\pi }(\text{sign }m)\{mv_{f}\text{
}\epsilon _{1}.\epsilon
_{2}+\frac{1}{4}\frac{1}{mv_{f}}(k_{2}.\epsilon _{1}\text{
}k_{1}.\epsilon _{2})\}  \label{35}
\end{equation}%
Here the divergence does depend on $m$ in addition to the $($sign
$m).$ In terms of electric and magnetic fields given above (time
independent $\vec{A}$
and uniform magnetic field) and neglecting the second term in Eq. (\ref{35}).%
\begin{equation}
\partial _{\mu }J_{3}^{\mu }=\frac{e^{2}}{4\pi }(\text{sign }m)mv_{f}\text{ }%
B^{2}r^{2}  \label{36}
\end{equation}%
One may take $m=\Delta /v_{f}^{2}$ where $\Delta$ is energy gap in
graphene's band structure, referred to as the Dirac gap \cite{11}.

\end{document}